\documentclass[12pt]{article}
\usepackage{amssymb}
\usepackage{amscd}

\begin{document}
\newcommand {\be}{\begin{equation}}
\newcommand {\ee}{\end{equation}}
\newcommand {\ba}{\begin{eqnarray}}
\newcommand {\ea}{\end{eqnarray}}
\newcommand {\bea}{\begin{array}}
\newcommand {\cl}{\centerline}
\newcommand {\eea}{\end{array}}
\renewcommand {\theequation}{\thesection.\arabic{equation}}
\renewcommand {\thefootnote}{\fnsymbol{footnote}}
\newcommand {\newsection}{\setcounter{equation}{0}\section}
\def \parth {{\partial\over\partial\theta_{\mu\nu}}}
\def \parmu {\partial_{\mu}}
\def \parnu {\partial_{\nu}}
\def \ola  {\overleftarrow}
\def \ora  {\overrightarrow}
\def \zz  {{\mathbb Z}}
\def \rr  {{\mathbb R}}

\def \com {commutative }
\def \ncy {noncommutativity }
\def \CS {Chern-Simons }
\def \nc {noncommutative }

\thispagestyle{empty}
\setcounter{page}0
 
\baselineskip 0.65 cm
\begin{flushright}
IC/2001/15 \\
hep-th/0103139
\end{flushright}
\begin{center}
{\Large{\bf Absence of Higher Order Corrections to\\ 
Noncommutative Chern-Simons Coupling}}
\vskip 1cm

{\bf {Ashok Das$^1$ and M.M. Sheikh-Jabbari$^2$}}

\vskip .5cm

$^1${\it Department of Physics and Astronomy\\
University of Rochester \\
Rochester, NY 14627-0171, USA}\\
{\tt das@pas.rochester.edu}\\
$^2${\it The Abdus Salam International Center for Theoretical Physics \\
Strada Costiera 11, Trieste, Italy}\\
{\tt jabbari@ictp.trieste.it}

\end{center}

\begin{abstract}
We analyze the structure of noncommutative pure Chern-Simons theory
systematically in the axial gauge. We show that there is no IR/UV
mixing in this theory in this gauge. In fact, we show, using the usual
BRST identities as well as the identities following from vector
supersymmetry, that this is a free theory. As a result, the tree level
Chern-Simons coefficient is not renormalized. It also holds that the
Chern-Simons coefficient is not modified at finite temperature. As a
byproduct of our analysis, we prove that the ghosts completely
decouple in the axial gauge in a noncommutative gauge theory.
\end{abstract}
\newpage
\newsection {Introduction}
The physics on \nc space (and space-time), after a new motivation from
string theory, has attracted a lot of interest \cite{SW}.
In addition to the string theory interests, \nc quantum mechanics
and \nc field theories have their own attractions
\cite{{Micu},{Chaich}} as well.
The \nc space can be defined by
\be\label{NC}
[\hat{x}^{\mu},\hat{x}^{\nu}]=i\theta^{\mu\nu}\ ,
\ee
where $\hat{x}^\mu$ are the space-time coordinate
operators and $\theta^{\mu\nu}$ is a constant anti-symmetric tensor.
In order to formulate \nc field theories, we need a realization of 
Eq. (\ref{NC}) in the space of the fields (functions). In fact such a
realization is simple to find, recalling the Weyl-Moyal correspondence
\cite{Micu}, according to which the \nc version of a given field theory
is obtained by replacing the usual product of functions by the $\star$-product:
\be\label{starp}
f(x)\star g(x)= f(x)\ {\rm exp}({i\over
2}\theta^{\mu\nu}\ola{\parmu}\;\ora{\parnu})g(x)\ .
\ee

It has been discussed that the \nc space appears naturally in planar
physics, namely, in $(2+1)$ dimensional theories. More specifically, the
\nc plane accommodates the physics of charged particles in a
(strong) magnetic field and, therefore, the quantum Hall physics
\cite{{Susskind},{Polych}}. 
In particular, it has been shown that the \nc version of pure \CS theory
captures the physics of Laughlin wave functions \cite{Susskind}, in
the sense that the level of the \nc $U(1)$ \CS theory plays the role of
the filling fraction. Being motivated by quantum Hall physics,
\nc \CS theories, both Abelian and non-Abelian, have been studied extensively 
\cite{{Bichl},{Wu},{Quant},{poly},{NCCS}}. 

The perturbative analysis and the one loop calculations of the \nc \CS
theories 
have, thus far, been performed in covariant gauges \cite{{Bichl},{Wu}}.
In the present work, following the lines of \cite{Das}, we will study the
\nc pure \CS in the axial gauge \cite{Leibbrandt} and present explicit
calculations at one loop. In the axial gauge, we have
the advantage of ghost decoupling as well as various cancellations. In
this way, we show, using
the BRST as well as the vector supersymmetry of \nc\CS theories, that the one
loop  result can be extended to all orders. 

The paper is organized as follows. In section {\bf 2}, we introduce the \nc 
\CS theory by giving its action and study the theory in the homogeneous
axial gauge. Then, we show the BRST invariance and the vector supersymmetry of
the gauge-fixed action. Using these symmetries we find the proper set of
Ward identities, which allows us to conclude that this is a free
theory. Explicit one loop calculations are presented in section {\bf
3} where we also show that ghosts decouple in the axial gauge in any
noncommutative gauge theory. The last section is devoted to some
concluding remarks. 

\section{Noncommutative Chern-Simons Theory}

In this section, we discuss various aspects of the noncommutative pure
Chern-Simons theory defined by
\begin{equation}
S_{\rm NCCS} = m \int_{M}
\epsilon^{\mu\nu\lambda}\,{\rm Tr}\left(A_{\mu}\star
\partial_{\nu}A_{\lambda} + {2ig\over 3} A_{\mu}\star A_{\nu}\star
A_{\lambda}\right)\ .
\label{A}\end{equation}
Here, the star product is defined in Eq. (\ref{starp}), $A_{\mu}$ takes
values in
the Lie algebra $u(N)$ (Hermitian $N\times N$ matrices) and \lq\lq Tr''
stands for the trace over the $u(N)$ indices in the 
representation where the generators satisfy ${\rm Tr}\,t^{A}t^{B} =
{1\over 2}\delta^{AB}$. We
assume that the base manifold $M$ can be separated
into a two dimensional (noncommutative) subspace times $R$ or $S^{1}$
and further assume that the $\star$-product (Eq. (\ref{starp})) can be defined
on the two dimensional part of this manifold.
The $\star$-product with constant $\theta$ can only be realized on flat
manifolds, namely, noncommutative plane, noncommutative
cylinder and the noncommutative torus. 
Here we present our calculations for the case of a \nc plane, although the
same arguments and results hold true also for \nc
cylinder and torus. 

The action in Eq. (\ref{A}) is invariant under the infinitesimal gauge
transformations
\begin{equation}
A_{\mu}\rightarrow A_{\mu} + \partial_{\mu}\lambda +
ig\left(\lambda\star A_{\mu} - A_{\mu}\star \lambda\right) =
D_{\mu}\lambda,\qquad \lambda\in u(N)\label{B}
\end{equation}
Under a finite gauge transformation, however, this action changes by a
constant (the Lagrangian density changes by a total derivative) and
the consistency of the theory then requires that the coefficient of
the \CS action should be quantized. 
We note that this result does not depend on $\theta$ and is also true 
for NC$U(1)$ case. In fact, the $\theta\to 0$ limit is not a smooth one.
This non-smoothness can be understood noting Eq. (\ref{NC}) and
the fact that,
in the \nc case (no matter how small $\theta$ is), we deal with
functions with countable number of degrees of freedom (for a proper
representation, see e.g. \cite{Harvey}), while in the \com case this is
uncountable.

The pure Chern-Simons theory is a
topological theory (both in the commutative and the
noncommutative cases) since the action does not depend on a
metric. However, a gauge fixing and the corresponding ghost actions
would depend on a metric and one might wonder whether the topological
nature of such theories will be destroyed by quantum corrections. In
the case of a commutative pure Chern-Simons theory, it is known that
the full theory including gauge fixing and ghosts is completely free
(when suitably regularized) \cite{Das}. In fact, this is why the theory is
topological. It is, therefore, interesting to ask if a
similar conclusion holds for the noncommutative pure Chern-Simons
theory. To analyze this question, let us recall that it is in the
axial gauge that the free nature of the commutative pure Chern-Simons
theory is best seen. In the noncommutative case, on the other hand,
there is a possibility of UV/IR mixing \cite{IRUV} and it is well known
that the
axial gauge is not particularly suitable in the presence of IR
divergences. Consequently, this question becomes even more interesting
and non-trivial in the present case.

The theory is defined in the homogeneous axial gauge $n^{\mu}A_{\mu} =
0$, where $n^{\mu}$ is a fixed direction in space, by adding a gauge
fixing and ghost action of the forms (the definition of the covariant
derivative is given in Eq. (\ref{B})and, for simplicity, we will
assume $n^{2}=1$)
\begin{equation}
S_{\rm gf} + S_{\rm ghost} = - \int_{M} {\rm Tr}\,\left({1\over \xi}(n\cdot
A)\star (n\cdot A) + \overline{c}\star (n\cdot Dc)\right)\label{C}
\end{equation}
to our theory and looking at the theory in the limit $\xi\rightarrow
0$. Alternately, one can define the gauge fixing and the ghost actions
with an auxiliary field (Lagrange multiplier field) of the form
\begin{equation}
S_{\rm gf} + S_{\rm ghost} = - \int_{M} {\rm Tr}\,\left(F\star (n\cdot
A) + \overline{c}\star (n\cdot Dc)\right)\ .\label{D}
\end{equation}
The Feynman rules can be read out from the theory
\begin{equation}
S = S_{\rm NCCS} + S_{\rm gf} + S_{\rm ghost}\label{E}
\end{equation}
We note that the gauge and the ghost propagators take the simple forms
\begin{eqnarray}
D_{\mu\nu}^{AB}(p) & = & {i\over m}\,\delta^{AB}
\epsilon_{\mu\nu\lambda}\,{n^{\lambda}\over (n\cdot p)}\ ,\nonumber\\
D^{AB}(p) & = & {i\delta^{AB}\over (n\cdot p)}\ .\label{F}
\end{eqnarray}
Here $A,B = 0,1,\cdots , N^{2}-1$ are the $u(N)$ indices, which is the 
same as the \com case (for more details, see \cite{Bonora}). It is
important to note that, in the homogeneous axial gauge, the gauge
propagator is transverse to $n^{\mu}$, namely,
\begin{equation}
n^{\mu}D_{\mu\nu}^{AB} (p) = 0\label{G}\ .
\end{equation}
The interaction vertices can also be read out quite easily and have
the forms
\begin{eqnarray}
{\rm Three\ gluon\ vertex:} \ && \Gamma_{\mu\nu\lambda}^{ABC}
(p_{1},p_{2},p_{3})  = mg\epsilon_{\mu\nu\lambda}\, Q_{\theta}^{ABC}
(p_{1},p_{2},p_{3}) \delta^{3}(p_{1}+p_{2}+p_{3})\ , \nonumber\\
{\rm Ghost-gluon\ vertex:}\  &&
\Gamma_{\mu}^{ABC} (p_{1},p_{2},p_{3}) =  gn^{\mu}\, Q_{\theta}^{ABC}
(p_{1},p_{2},p_{3})\delta^{3}(p_{1}+p_{2}+p_{3}),\label{H}
\end{eqnarray}
where
\begin{equation}
Q_{\theta}^{ABC} (p_{1},p_{2},p_{3}) = (f^{ABC} \cos {p_{1}\theta p_{2}\over 2} - d^{ABC}
\sin {p_{1}\theta p_{2}\over 2})\ .\label{I}
\end{equation}
For $f^{ABC}$ and $d^{ABC}$ factors and their normalizations we use the
conventions of \cite{Bonora}, i.e. if we denote the generators on
$u(N)$ by $t^A$, then, 
\be
t^A t^B={i\over 2}f^{ABC}t^C+{1\over 2} d^{ABC}t^C\ ,
\ee
where $f^{ABC}$ is totally anti-symmetry and $d^{ABC}$ completely
symmetric. Also, $t^0={1\over\sqrt{2N}}{\bf 1}$ and $f^{0AB}=0$ and
$d^{0AB}=\sqrt{{2\over N}}\delta^{AB}$.

It is immediately clear from the Feynman rules of the theory that
because the ghost interaction vertex involves a factor of $n^{\mu}$,
and that the gauge propagator is transverse to this direction,
there can be no quantum correction to the ghost two point and three
point functions. In other words, the wave function and the vertex
renormalization for the ghosts is trivial, $\tilde{Z}_{1} = 1 =
\tilde{Z}_{3}$. This also happens in the commutative gauge theories in the
axial gauge. In fact, in the commutative theories, one can show that
the ghosts decouple completely from the theory in the axial
gauge. Here, we see that the ghosts decouple, at least in diagrams
involving open ghost lines. It is not {\it a priori} clear that the
ghosts would decouple in diagrams involving closed loops. We will come
to this issue shortly.

\subsection{BRST symmetry}

Even though gauge fixing and ghost Lagrangians break the $u(N)$ gauge
invariance of Eq. (\ref{B}), the full theory is invariant under the
fermionic BRST symmetry
\begin{eqnarray}
\delta A_{\mu} & = & \omega (D_{\mu}c)\nonumber\\
\delta c^{A} & = & ig\omega c\star c\nonumber\\
\delta \overline{c} & = & - \omega F\nonumber\\
\delta F & = & 0\label{J}
\end{eqnarray}
Here, $\omega$ is a constant anti-commuting parameter of
transformation and the invariance of the total action is easily seen
in the formulation with the auxiliary field. It is in this
formulation also that the BRST symmetry is nilpotent off-shell. One
can now derive the Ward identities (Slavnov-Taylor identities) of the
theory following from the BRST invariance in a standard
manner. Without going into details, we simply note that the Ward
identities take the form 
\begin{equation}
\int_{M} \left({\delta\Gamma\over \delta A_{\mu}}\star
{\delta\Gamma\over \delta K^{\mu}} - {\delta\Gamma\over \delta c}\star
{\delta\Gamma\over \delta K} + F\star {\delta\Gamma\over \delta
\overline{c}}\right) = 0\ ,\label{J1}
\end{equation}
where $K^{\mu}$ and $K$ are sources for the composite
BRST variations $D_{\mu}c$ and $ic\star c$ respectively. This form of
the Slavnov-Taylor identity holds in any gauge. However, it leads to very
interesting consequences in the axial gauge. For example, as we have
already noted, in the axial gauge ghosts decouple, at least in
diagrams involving open ghost lines because of the structure of the
ghost interaction as well as Eq. (\ref{G}). As a result, the terms in
the effective action involving the sources $K^{\mu}$ and $K$ do not
renormalize either, much like the ghost fields and  the ghost
interaction vertex. An immediate consequence of this is the fact that, in the
axial gauge, the wave function and the vertex renormalization for the
gauge fields are equal, namely, $Z_{1} = Z_{3}$. This does not,
however, imply that they are trivial.

It is worth pointing out here that, in the axial gauge, the complete two point
function can be parameterized, in general, as
\begin{equation}
\Gamma_{\mu\nu}^{AB} (p) = \delta^{AB}\left(im
\epsilon_{\mu\nu\lambda} p^{\lambda} (1 +\Pi_{2}(p^{2})) + (p^{\mu} -
{p^{2}n^{\mu}\over n\cdot p})(p^{\nu}-{p^{2}n^{\nu}\over n\cdot p})
\Pi_{3}(p^{2})\right)
\end{equation}
where $\Pi_{2}$ and $\Pi_{3}$ represent quantum corrections to the two
tensor structures. 
This implies that we can determine the \CS coefficient (including the
quantum corrections) from the complete two point function as
\begin{equation}
m (1+\Pi_{2}(0))\,\delta^{AB} = - {i\over 6}\epsilon^{\mu\nu\lambda}
\left.{\partial\Gamma_{\mu\nu}^{AB}(p)\over \partial p^{\lambda}}\right|_{p=0}
\end{equation}

\subsection{Vector supersymmetry}

In addition to the BRST invariance, the total action possesses another
fermionic symmetry, namely, under
\begin{eqnarray}
\delta A_{\mu} & = & \epsilon_{\mu\nu\lambda}
\epsilon^{\nu}n^{\lambda}c\nonumber\\
\delta c & = & 0\nonumber\\
\delta \overline{c} & = & - \epsilon^{\mu} A_{\mu}\nonumber\\
\delta F & = & \epsilon^{\mu}\partial_{\mu}c\label{K}
\end{eqnarray}
the total action can be easily checked to be invariant. Here,
$\epsilon^{\mu}$ is a constant anti-commuting vector parameter. It is
interesting to note that, unlike the BRST symmetry, these symmetry
transformations are linear in the field variables. Consequently, the
corresponding Ward identities can be derived without the need for any
extra sources (there are no composite variations) and take the form
\begin{equation}
\int_{M} \left(\epsilon^{\mu\nu\lambda} {\delta\Gamma\over \delta
A^{\nu}}\star c\, n^{\lambda} - {\delta\Gamma\over \delta F}\star
\partial^{\mu}c + A^{\mu}\star {\delta\Gamma\over \delta
\overline{c}}\right) = 0\label{L}
\end{equation}
While the usual Slavnov-Taylor identities in Eq. (\ref{J1}) only imply
$Z_{1}=Z_{3}$ (and nothing about the triviality of these quantities)
in the axial gauge, the identities following from the vector
supersymmetry lead to the fact that this is a free theory. For example,
differentiating Eq. (\ref{L}) with respect to ${\delta^{2}\over \delta
A^{\mu}\delta c}$ and setting all fields equal to zero, we obtain (in
momentum space),
\begin{equation}
\epsilon^{\mu\nu\lambda}\,{\delta^{2}\Gamma\over \delta
A^{\mu}(-p)\delta A^{\nu}(p)}\,n_{\lambda} +
ip^{\mu}\,{\delta^{2}\Gamma\over \delta A^{\mu}(-p)\delta F(p)} -
3\,{\delta^{2}\Gamma\over \delta\overline{c}(-p)\delta c(p)} =
0\label{L1}
\end{equation}
Since neither the ghost two point function nor the mixing of the
auxiliary field with the gauge field are renormalized in the axial
gauge, it immediately follows that the two point function for the
gauge field  is not renormalized for this theory either. Similarly,
one can show from Eq. (\ref{L}) (taking higher derivatives) that the
gauge three point vertex, in this theory, is not renormalized as
well, i.e. the theory is a {\it free theory}.

Thus, formally, we have shown that, in the axial gauge, the
noncommutative pure
Chern-Simons theory is a free theory and, consequently, the \CS
coefficient is not renormalized at all. All of this, of
course, depends heavily on the identities following from the vector
supersymmetry invariance of the theory. Hence, such a conclusion will
hold as long as we regularize the theory maintaining this
symmetry. If, on the other hand, the
regularization breaks this symmetry, such a conclusion will not hold
and one may have anomalous contributions. It is worth pointing out here
that, in literature, the pure Chern-Simons theory has been considered
as the heavy mass limit of the Yang-Mills-Chern-Simons theory (namely,
$m\rightarrow\infty$) \cite{{Wu},{Deser}}. In this case, one can think of
the Yang-Mills
theory as providing a regularization. However, the Yang-Mills theory,
itself, 
is not invariant under the vector supersymmetry transformations of
Eq. (\ref{K}) (the same holds true in the Landau gauge as
well.). Consequently, regularizing the theory in this manner violates
the vector supersymmetry invariance and we would not expect our
conclusion on the free nature of the theory to hold in this case. In
fact, as is well known, the
calculations from the Yang-Mills-Chern-Simons theory in the limit
$m\rightarrow \infty$ show a shift of the \CS coefficient at the one
loop level. However, this shift is a quantized one in both \nc and \com
theories \cite{{Quant},{Das},{Rao},{Martin}}.

\section{One-Loop Calculations}

The arguments of the last section on the free nature of the \nc pure
Chern-Simons theory were based on algebraic identities and were quite
formal. It is, therefore, essential to check these through an explicit
one loop calculation. This is also important from another point of
view. Namely, in the axial gauge in commutative theories, one knows
that a consistent prescription for handling unconventional poles is
the principal value prescription. It is not clear whether such a
prescription will continue to hold in the noncommutative
case. Furthermore, an explicit calculation is likely to shade light on
the question of ghost decoupling in closed loops in this theory.

The calculation of the self-energy, at one loop, is trivial. From the
Feynman rules in Eqs. (\ref{F},\ref{H},\ref{I}), one can write down
the ghost and the gauge contribution to the gauge self-energy as (all
momenta are defined as incoming and we are ignoring factors of
$(2\pi)^{3}$ in the integrals)
\begin{eqnarray}
\Pi_{\mu\nu}^{(1)\,AB (ghost)}(p) & = & g^{2}n^{\mu}n^{\nu} \int
d^{3}k\,{Q_{\theta}^{ACD} (p,-(k+p),k) Q_{\theta}^{BDC}
(-p,-k,k+p)\over (n\cdot k)(n\cdot (k+p))}\nonumber\\
\Pi_{\mu\nu}^{(1)\,AB (gauge)}(p) & = & - g^{2}n^{\mu}n^{\nu} \int
d^{3}k\,{Q_{\theta}^{ACD} (p,-(k+p),k) Q_{\theta}^{BDC}
(-p,-k,k+p)\over (n\cdot k)(n\cdot (k+p))}\label{L2}
\end{eqnarray}
The two contributions cancel each other and, consequently,
the correction to the self-energy identically vanishes,
\begin{equation}
\Pi_{\mu\nu}^{(1)\,AB}(p) = \Pi_{\mu\nu}^{(1)\,AB (ghost)}(p) +
\Pi_{\mu\nu}^{(1)\,(gauge)}(p) = 0
\end{equation}
This is completely consistent with our conclusions in the last
section. Without giving details, we simply comment here that each
of the two integrals in (\ref{L2}) can be evaluated (see
discussion below for the triangle graph) and be shown to vanish
individually. Thus, even if
the ghost and the gauge loop corrections to self-energy did not cancel
each other, their individual contributions would have been trivial. This is
important because it shows that the ghosts do not contribute in the
closed self-energy loop. On the other hand, the fact that the two
contributions identically cancel each other has important consequence
at finite temperature, as we will argue in the last section.

In analyzing the correction to the gauge interaction vertex, we note
that, in the axial gauge, the usual Slavnov-Taylor identity,
$Z_{1}=Z_{3}$, would imply that there should be no correction to the
triangle graph as well. However, if we naively write down the ghost
and the gauge contributions to the triangle graph, they take the forms
(One has to add the contribution from the crossed graph as well, which
we are neglecting for the moment.)
\begin{eqnarray}
&  & \Gamma_{\mu\nu\lambda}^{(1)\,ABC(ghost)}(p_{1},p_{2},-(p_{1}+p_{2}))=
\nonumber\\
& = &
ig^{3} n_{\mu}n_{\nu}n_{\lambda} \int
d^{3}k\, Q_{\theta}^{AA'C'}(p_{1},-(k+p_{1}),k)
Q_{\theta}^{BB'A'}(p_{2},-(k+p_{1}+p_{2}),k+p_{1})\nonumber\\
&  & \times 
{Q_{\theta}^{CC'B'}(-(p_{1}+p_{2}), -k,k+p_{1}+p_{2}) \over (n\cdot
k)(n\cdot (k+p_{1}))(n\cdot (k+p_{1}+p_{2}))}\nonumber\\
&  & \Gamma_{\mu\nu\lambda}^{(1)\,ABC
 (gauge)}(p_{1},p_{2},-(p_{1}+p_{2}))\nonumber\\
 & = & -2ig^{3} n_{\mu}n_{\nu}n_{\lambda} \int
d^{3}k\,Q_{\theta}^{AA'C'}(p_{1},-(k+p_{1}),k)
Q_{\theta}^{BB'A'}(p_{2},-(k+p_{1}+p_{2}),k+p_{1})\nonumber\\
 &  & \times
{Q_{\theta}^{CC'B'}(-(p_{1}+p_{2}), -k,k+p_{1}+p_{2})\over (n\cdot
k)(n\cdot (k+p_{1}))(n\cdot (k+p_{1}+p_{2}))}\label{M}
\end{eqnarray}
Therefore, the ghost and the gauge contributions do not seem to cancel
each other in the triangle graph. Let us note
that these graphs are logarithmic divergent by naive power counting and,
therefore, if they do not vanish, then this would require adding
counter terms that explicitly depend on $n^{\mu}$. Such counter terms
are known to exist in a general axial gauge \cite{Leibbrandt}, but do
not arise  in the
homogeneous axial gauge (in commutative theories). Moreover, the
triangle diagram, if it does not vanish would violate the
Slavnov-Taylor identities of the theory. Therefore, we analyze the
structure of these graphs carefully in some detail.  

Let us recall that in the commutative gauge theories (in axial gauge)
integrals of the kind
\[
\int d^{n}k\,{1\over (n\cdot k)\cdots (n\cdot (k+p_{1}+\cdots
+p_{m-1}))}
\]
can all be regularized to zero using dimensional regularization
\cite{Frenkel}. (This
is, in fact, one of the ways to see that ghosts decouple in the axial
gauge.) It is
not, however, clear whether this holds in noncommutative gauge
theories, where the interaction vertices involve nontrivial phase
factors. Let us, therefore, analyze the basic integral of the kind
\begin{equation}
I = \int d^{3}k\,{e^{ik\cdot p}\over (n\cdot k)(n\cdot
(k+p_{1}))(n\cdot (k+p_{1}+p_{2}))}\label{M1}
\end{equation}
where $p^{\mu}$ is an arbitrary, external variable. First, we note
that combining fractions, we can write the denominator of the
integrand also as
\begin{eqnarray}
 &  & {1\over (n\cdot p_{1})(n\cdot (p_{1}+p_{2}))(n\cdot k)} -
 {1\over (n\cdot p_{1})(n\cdot p_{2}(n\cdot (k+p_{1}))}\nonumber\\
 &  & \quad + {1\over (n\cdot p_{2}) (n\cdot (p_{1}+p_{2}))(n\cdot
 (k+p_{1}+p_{2}))}\label{M2}
\end{eqnarray}
Using Eq. (\ref{M2}) in (\ref{M1}) and shifting variables of
integration, we obtain
\begin{equation}
I = \left[{1\over (n\cdot p_{1})(n\cdot (p_{1}+p_{2}))} - {e^{-ip_{1}\cdot
p}\over (n\cdot p_{1})(n\cdot p_{2})} + {e^{-(p_{1}+p_{2})\cdot
p}\over (n\cdot p_{2})(n\cdot
(p_{1}+p_{2}))}\right]\,\bar{I}\label{M3}
\end{equation}
where
\begin{equation}
\bar{I} = \int d^{3}k\, {e^{ik\cdot p}\over (n\cdot k)}
\end{equation}
which can be evaluated using the principal value prescription 
(standard in axial gauges) \cite{Frenkel1} to give
\begin{equation}
\bar{I}^{PV} = \delta^{2}(p^{T}) \int_{-\infty}^{\infty}
d\omega\,{e^{i\omega (n\cdot p)}\over \omega} = i\pi sgn(n\cdot
p)\delta^{2}(p^{T})\label{M4}
\end{equation}
Here, we have decomposed $p^{\mu}$ into longitudinal and transverse
directions with respect to $n^{\mu}$ and $p^{\mu T}$ represents the
transverse component. We could also have used the Feynman prescription
to evaluate the integral, which would have given
\begin{equation}
\bar{I}^{F} = -2i\pi \theta(-n\cdot p) \delta^{2}(p^{T})
\end{equation}
where $\theta$ represents the conventional step function (and not the
parameter of noncommutativity). There are several things to note here. First of
all, unlike in the commutative theories (when there is no phase), such
an integral is not automatically zero. Second, we note that, even in
the presence of a phase factor, such an integral has a finite value
whether we use the principal value prescription or the Feynman
prescription for the pole, implying that there is no IR/UV mixing. In
the following discussion, we will use the principal value
prescription, but our conclusion, as will become clear, holds for the
other case as well. 

Using Eq. (\ref{M4}) in (\ref{M3}), we can write
\begin{eqnarray}
 &  & \int d^{3}k\,{\cos k\cdot p\over (n\cdot k)(n\cdot
 (k+p_{1}))(n\cdot (k+p_{1}+p_{2}))}\nonumber\\
& = & -\pi sgn(n\cdot p)\delta^{2}(p^{T})\left[{\sin p_{1}\cdot p\over
 (n\cdot p_{1})(n\cdot p_{2})} - {\sin (p_{1}+p_{2})\cdot p\over
 (n\cdot p_{2})(n\cdot (p_{1}+p_{2}))}\right]\nonumber\\
 &  & \int d^{3}k\,{\sin k\cdot p\over (n\cdot k)(n\cdot
 (k+p_{1}))(n\cdot (k+p_{1}+p_{2}))}\label{M5}\\
&= &\!\! \pi sgn(n\cdot p)\delta^{2}(p^{T})\left[{1\over (n\cdot
 p_{1})(n\cdot (p_{1}+p_{2}))} - {\cos p_{1}\cdot p\over (n\cdot
 p_{1})(n\cdot p_{2})} + {\cos (p_{1}+p_{2})\cdot p\over (n\cdot
 p_{2})(n\cdot (p_{1}+p_{2}))}\right]\nonumber
\end{eqnarray}
These are the two basic integrals we will need to analyze the triangle
graph. With these, let us go back to the triangle graph. Each of the
integrands involves four possible group theoretic structures and using
Eq. (\ref{M5}) as well as simple trigonometric identities. It is easy
to check that terms with an odd number of factors of $d^{ABC}$
(namely, three $d^{ABC}$ or two $f^{ABC}$ and one $d^{ABC}$)
identically vanish. On the other hand, terms with an even number of
factors of $d^{ABC}$ (no $d^{ABC}$ or one $f^{ABC}$ and two $d^{ABC}$)
do not. It is worth noting that, as we expect, there are \lq\lq planar" and 
\lq\lq non-planar" parts, in the sense that after some trigonometric
gymnastics, 
we find two generic types of integrals. One type has  the $\theta$
dependent phase factor not involving the loop-momentum, 
these are the planar parts. Essentially the form of the planar loop integrals
are the same as the \com case and hence they vanish after regularization.
The second type has the \nc phase factor dependent on loop-momentum
$k$,  and this is the non-planar part. These are the integrals we are concerned
with here. The terms with three factors of $f^{ABC}$ can be evaluated
using Eq. (\ref{M5}) and have the form (neglecting an overall factor)
\begin{eqnarray}
I_{1}^{ABC} & = & {\pi N\over 4}\,f^{ABC} \sin {p_{1}\theta p_{2}\over
2}\left[{sgn(n\cdot \theta p_{1})\delta^{2}((\theta p_{1})^{T})\over
(n\cdot p_{2})(n\cdot (p_{1}+p_{2}))} + {sgn(n\cdot \theta p_{2})\over
(n\cdot p_{1})(n\cdot (p_{1}+p_{2}))}\right.\nonumber\\
 &  & \quad \left. + {sgn(n\cdot \theta
(p_{1}+p_{2})\delta^{2}((\theta (p_{1}+p_{2}))^{T})\over (n\cdot
p_{1})(n\cdot p_{2})}\right]\label{M6}
\end{eqnarray}
The terms involving one factor of $f$ and two $d$'s can similarly be
evaluated to give
\begin{eqnarray}
I_{2}^{ABC} & = & -{\pi N\over 4}\,f^{ABC} \sin {p_{1}\theta p_{2}\over
2}\left[{sgn(n\cdot \theta p_{1})\delta^{2}((\theta p_{1})^{T})\over
(n\cdot p_{2})(n\cdot (p_{1}+p_{2}))} + {sgn(n\cdot \theta p_{2})\over
(n\cdot p_{1})(n\cdot (p_{1}+p_{2}))}\right.\nonumber\\
 &  & \quad \left. + {sgn(n\cdot \theta
(p_{1}+p_{2})\delta^{2}((\theta (p_{1}+p_{2}))^{T})\over (n\cdot
p_{1})(n\cdot p_{2})}\right]\label{M7}
\end{eqnarray}
Here, we have used various group theoretic identities (given in 
 \cite{Bonora}) to simplify the
final structure. It is clear now, that even though the individual
contributions to the triangle graph of the structures $fff$ and $fdd$
are  finite and nonzero, the sum
identically vanishes.
\begin{equation}
I_{1}^{ABC} + I_{2}^{ABC} = 0
\end{equation}
In other words, each of the two integrands in
Eqs. (\ref{M}) individually vanishes and, consequently, there is no
correction to the gauge interaction vertex. This is, of course, exactly what
we had concluded from the identities following from vector
supersymmetry. More importantly, though, it shows that the ghosts in
closed loops decouple in this theory, both in the self-energy and the
triangle graphs. With these results, one can show from Eq. (\ref{J1}),
with a little bit of analysis, that in this theory, for $n\geq 4$,
\begin{equation}
p_{n,\mu_{n}}\Gamma^{\mu_{1},\cdots ,\mu_{n}}_{A_{1},\cdots ,A_{n}}
(p_{1},\cdots ,p_{n}) = 0\label{N'}
\end{equation}
which is true for every external momentum. The $n$-point function, of course, gets
contributions from a ghost loop as well as a gauge loop. Both of them
have the same structure (as we have already seen in the case of the
triangle graph), but different coefficients (the different coefficient
for the gauge loop arises primarily because of the contractions of the
$\epsilon^{\mu\nu\lambda}$ tensors). Therefore, they cannot cancel
each other. Rather, the ghost loop as well as the gauge loop have to
be individually transverse to the external momenta. However, each of
these graphs is proportional to $n^{\mu_{1},\cdots ,\mu_{n}}$ and,
therefore, can be transverse to the external momenta provided it is
proportional to $\delta(n\cdot p_{1})\cdots\delta(n\cdot p_{n})$. On
the other hand, by the method of combining fractions, we can show that
each of these graphs is proportional to $\bar{I}$ (multiplied by
factors of trigonometric functions of external momenta) which we have
already  evaluated
and which does not have this property. Consequently, it follows that,
for Eq. (\ref{N'}) to hold, each
of the graphs must vanish independently. This, therefore, shows that
the theory is truly free. More than that it shows that ghosts in
closed loops do not contribute. With our earlier result that
ghosts in open lines do not contribute in the axial gauge, it is clear
that, in this theory, ghosts completely decouple. Furthermore, since
the structure of the ghost propagator and the ghost interaction vertex
in the axial gauge is independent of the nature of the gauge theory we
are considering (we may have a more complicated theory such as a
noncommutative Yang-Mills-Chern-Simons theory or a noncommutative
Yang-Mills theory, for example), the ghost loop will
vanish independent of the theory we are considering. Therefore, we
have proved the more general result that the ghosts completely
decouple in the axial gauge in a noncommutative gauge theory.

\section{Conclusion}

In this paper, we have studied systematically the structure of \nc pure
Chern-Simons theory in the homogeneous axial gauge. There are many
interesting results that one finds in this theory. First, of all, we
have shown that there is no problem in using the homogeneous axial
gauge in such theories and the principal value prescription makes
individual integrals well behaved. At one loop, we have shown that
there is no quantum correction, which is consistent with the conclusion that
this is a free theory. Such a conclusion follows from the identities
coming from the vector supersymmetry of the theory together with the usual
Slavnov-Taylor identities. Our one loop calculation shows that there
is no IR/UV mixing and that the Slavnov-Taylor identities hold in such
theories in the axial gauge. In fact, the quantum behavior of this \nc
theory is very much like its commutative counterpart \cite{Das}. Without going
into  details, let us simply note here that, since Slavnov-Taylor
identities hold in the axial gauge in this theory, by a generalization of the
arguments in \cite{Das}, it can also be shown, in the \nc
Yang-Mills-Chern-Simons theory, that there is no correction to the \CS
coefficient beyond one loop in the axial gauge. Another way of saying
this is that the ratio ${4\pi m\over g^{2}}$ receives no correction
beyond one loop in any gauge. This is quite important from the point
of view of the consistency of the theory under {\it large gauge}
transformations. Finally, let us note that it has been proposed to use
the \nc pure Chern-Simons theory to describe quantum Hall effect. In
this connection, the behavior of the \CS coefficient at finite
temperature is quite crucial, if it were to describe a phase
transition. We note that, at finite temperature, gauge fields and
ghosts obey the {\it same statistics} \cite{Das1} so that the finite
temperature propagators, in the axial gauge (in real time formalism),
have  the forms
\begin{eqnarray}
D_{\mu\nu}^{AB\,(\beta)}(p) & = & {i\delta^{AB}\over
m}\epsilon_{\mu\nu\lambda} n^{\lambda}\left({1\over n\cdot p} - 2i\pi
n_{B}(|p^{0}|)\delta (n\cdot p)\right)\nonumber\\
D^{AB\,(\beta)}(p) & = & i\delta^{AB}\left({1\over n\cdot p} - 2i\pi
n_{B}(|p^{0}|)\delta (n\cdot p)\right)\label{N}
\end{eqnarray}
Since there is no modification of the interaction vertices at finite
temperature, from the forms of Eqs. (\ref{L2}) and (\ref{N}), it is
clear that the correction to the self-energy will vanish even at
finite temperature in the \nc pure Chern-Simons theory. As a result,
the \CS coefficient cannot have a temperature dependence and,
consequently, it would appear that such a model may not have a
satisfactory behavior to describe quantum Hall phase transitions. More
precisely, the quantum Hall fluid-Wigner crystal phase transition cannot be
driven by temperature effects. So, it seems that the \nc \CS approximation
is not enough for describing this phase transition.

\vskip 1cm
We would like to thank the organizers of the $37^{th}$ Karpacz Winter
School, where this work was started, for hospitality and a warm atmosphere.
One of us (A.D.) would like to thank Profs. J. Frenkel and
J. C. Taylor for many helpful discussions, particularly, in
connection with the evaluation of the axial gauge integrals.
This work was supported in part by US DOE Grant No. DE-FG-02-91ER40685.

\end{document}